\documentclass[onecolumn,10pt,cleanfoot]{asme2ej}

\usepackage{graphicx}
\usepackage{amssymb}
\usepackage{amsmath}
\usepackage[ruled,vlined]{algorithm2e}
\usepackage[noend]{algpseudocode}
\usepackage{bbold}
\usepackage{bm} 
\usepackage[version=4]{mhchem}
\usepackage{siunitx}
\usepackage{longtable,tabularx}
\usepackage{subcaption}
\usepackage{float}
\usepackage{bm}
\usepackage{ragged2e}
\justifying
\usepackage{amsmath}
\usepackage{xcolor}
\usepackage{multicol}
\usepackage{url}

\newcommand{\mtx}[0]{\mathbf}  
\newcommand{\vc}[0]{\boldsymbol}  

\newcommand{\scriptN}[0]{\mathcal N}

\newcommand{\bY}{\mathbf{Y}}

\newcommand{\btheta}{\boldsymbol{\theta}}
\newcommand{\bphi}{\boldsymbol{\phi}}
\newcommand{\highlighttext}[1] {\textcolor{black}{#1}}

\title{Inverse Aerodynamic Design of Gas Turbine Blades using Probabilistic Machine Learning }

\author{Sayan Ghosh\thanks{Address all correspondence to this author.} $^{1}$, 
	Govinda Anantha Padmanabha$^{2}$, 
	Cheng Peng$^{2}$,
	Valeria Andreoli$^{1}$, 
	Steven Atkinson$^{1}$,\\
	\textbf{Piyush Pandita$^{1}$,
	Thomas Vandeputte$^{1}$,
	Nicholas Zabaras$^{2}$,
	Liping Wang$^{1}$} 
    \affiliation{\\
	$^{1}$General Electric Research\\
	Niskayuna, New York, 12309\\ \\

    $^{2}$Department of Aerospace and Mechanical Engineering\\
    University of Notre Dame\\
    Notre Dame, Indiana 46556 \\ \\
    
    Email: sayan.ghosh1@ge.com
    
    }	
}

\begin{document}

\maketitle    

\begin{abstract}
{\it One of the critical components in Industrial Gas Turbines (IGT) is the turbine blade. 
Design of turbine blades needs to consider multiple aspects like aerodynamic efficiency, durability, safety and manufacturing, which make the design process sequential and iterative.
The sequential nature of these iterations forces a long design cycle time, ranging from several months to years. 
Due to the reactionary nature of these iterations, little effort has been made to accumulate data in a manner that allows for deep exploration and understanding of the total design space. 
This is exemplified in the process of designing the individual components of the IGT resulting in a potential unrealized efficiency.
To overcome the aforementioned challenges, we demonstrate a probabilistic inverse design machine learning framework (PMI), to carry out an explicit inverse design.
PMI calculates the design explicitly without excessive costly iteration and overcomes the challenges associated with ill-posed inverse problems. 
In this work, the framework will be demonstrated on inverse aerodynamic design of three-dimensional turbine blades. }

\textbf{Keywords:} Inverse design, Invertible Neural Network, Machine Learning, Probabilistic Modeling, Gaussian Process, Gas Turbine, Aerodynamic Design
\end{abstract}

\section{INTRODUCTION}

The design of industrial gas turbines (IGT) is a multi-disciplinary, complex process which often requires numerous iterations across different teams to achieve multiple design objectives. 
Every subsystem of the turbine must be designed considering various multi-disciplinary targets like performance, manufacturability, temperature limits, component life, simplicity of assembling and disassembling procedures. 
Often, design objectives and constraints conflict and numerous iterations at system level (firing temperature, pressure ratio, engine flow, etc.), module level (compressor, combustor, turbine, etc.), and component level (individual airfoils) are required in order to achieve a cross-functionally valid design. 
Many of these iterations occur serially, starting from an aerodynamic assessment, followed by heat transfer and mechanical assessments with several internal design loops within and between each discipline. 
The sequential nature of these iterations forces long design cycle times, ranging from several months to years.

An inverse design process aims to positively impact the difficulties that encumber the traditional iterative process. 
In inverse design, the performance target is set as an input to the process, while the design parameters become the output. 
Using this approach, a design is explicitly generated from target performance metrics. 
Over the last few decades, various approaches of inverse design have been developed for a range of applications. 
These methods can be broadly classified into two categories: a) an optimization-based approach, and b) a  direct inverse design approach \cite{bui2004aerodynamic,dulikravich1999aerodynamic,obayashi1996genetic,dulikravich1995shape,bell1991inverse}. 
The former is similar to forward design methods, whereas the latter is used to find a design that satisfies all constraints and achieves the target performance. 
Most of the methods used for turbomachinery inverse designs are CFD-based algorithms, where the flow fields information is used to guide the blade redesign to match a given design target \cite{boselli2011inverse, deVito2003, Borges1990}. 
Recent work on inverse design has focused on developing frameworks that allow the user to quantify uncertainty around the optimal user inputs or design parameters for multiple applications, see \cite{ nellippallil2017goal,nickless2018comparison,white2021efficient,hou2020data}.
Specifically, work treating problems in the turbomachinery domain includes the work of \cite{bonaiuti2009coupling,roy2018inverse}, where the frameworks proposed leverage state-of-the-art optimization techniques to solve for the optimal input. The basic gap is the inability of the frameworks to combine and generate data from multiple sources under a limited budget and explicit inverse mapping that allows one to generate "x" as a function of "y".

With recent advances in the area of \emph{deep learning}, new methods have revolutionized the inverse design process \cite{jin2017deep,chen2017low,adler2017solving,adler2018learned,hauptmann2018model,mosser2020stochastic,lunz2018adversarial,laloy2018training}. 
However, the current methods struggle with high-dimensional design spaces, require large amounts of high-fidelity training data, and have difficulty dealing with problems that are ill-posed or ill-conditioned. 

In this work, we demonstrate a scalable framework for explicit inverse design, named Probabilistic Machine Learning for Inverse Design of Aerodynamic Systems (PMI)  \cite{ghosh2021pro}, that overcomes these challenges to enable inverse aerodynamic design of turbine blade airfoils. 
Given some desired performance quantities, the inverse problem is solved explicitly by sampling from the PMI’s explicit inverse mapping, making it trivial to characterize the marginal posterior density over potential designs which captures both the fundamental ill-posed nature of the inverse problem and the reality of epistemic uncertainty induced by limited resources to query the underlying physics of the system. 

The designs of interest in this work are turbomachinery components that are applicable to not only IGTs, but also to aviation turbine engines, wind turbines, and hydro turbines. 
As a representative problem that demonstrates the capabilities of the PMI framework, we will focus on the inverse design of the Last Stage Blade (LSB) cross-sectional airfoils.
As the largest rotating component in the IGT, the LSB is one of the most mechanically challenging components to design and often determines the total power output of the machine. 
The increasing push for larger and hotter turbines to drive down the cost of electricity has resulted in increasingly challenging LSB designs.
With deference typically skewed towards durability, this has generally resulted in greater aerodynamic compromises that negatively impact blade efficiency. 
This also drives longer design cycles as designers incrementally search for acceptable aero-mechanical solutions. 
Methods that enable reduction in design cycle time while also improving aerodynamic efficiency are therefore highly desirable.
In the next section, we will give brief details of PMI framework. 
The following section, includes the demonstration of our PMI framework on a toy problem. 
In Sec. 4, we demonstrate the framework on inverse aerodynamic design of 3D blade of last stage blade of IGT.
Finally, we will present conclusions and discussion in Sec. 5.

\begin{figure*}[t]
	\begin{center}
		\includegraphics[scale=0.45]{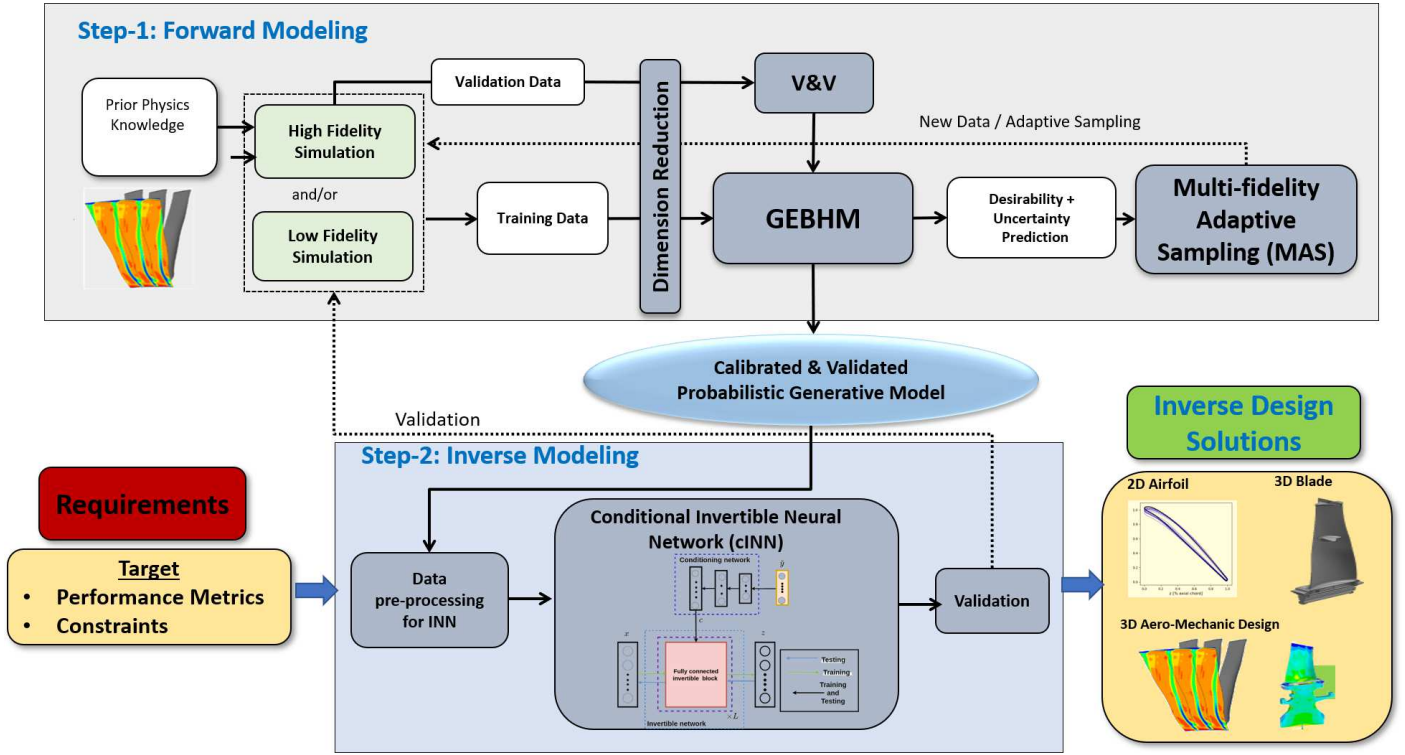}
	\end{center}
	\caption{Probabilistic Machine Learning for Inverse Design of Aerodynamic Systems (PMI) Framework}
	\label{fig:proml_ideas} 
\end{figure*}

\section{PMI FRAMEWORK}
The PMI framework, as shown in Figure \ref{fig:proml_ideas}, generates an explicit functional representation of the inverse design process. 
The main element of this framework is a conditional invertible neural network (cINN). 
The cINN is used to represent designs explicitly for the desired targets of performance and constraints. 
The framework entails the following two-step process: 
\begin{enumerate}
    \item Forward Modeling: efficient modeling of the forward process
    \item Inverse Modeling: training a conditional invertible neural network (cINN) using the data generated in the first step.
\end{enumerate}
In the forward modeling step, a probabilistic multi-fidelity Gaussian Process (MFGP) regression model for the expensive experiments is constructed using the GE Bayesian Hybrid Modeling (GEBHM) \cite{ghosh2020advances,pandita2021scalable}.
To reduce the cost associated with the design of the computer experiments~\cite{ghosh2019strategy,pandita2016extending,pandita2018stochastic,pandita2021surrogate,hennig2012entropy} required by the GEBHM, a multi-fidelity adaptive sampling \cite{ghosh2019strategy} is used to adaptively determine the experiment and level of fidelity that are needed to enhance the performance. 
The data generated in step 1 using the surrogate of the forward model (MFGP) will be used to train the cINN in step 2. 

In the inverse modeling step, a cINN is trained, which is a parameterized Bayesian bijective transform between the input or the design variables and the performance characteristics or the engineering quantities of interest (QoI).
The invertible block of cINN is built using a flow-based conditional generative model.
One of the key characteristics of the architecture used is its ability to capture the uncertainty due to the forward process as well as the epistemic uncertainty due to limited data, allowing engineers to carry out a trade-off between cost and risk.
To solve an inverse problem for a given target performance or QoI, the cINN’s explicit inverse mapping can be trivially sampled to generate potential designs while capturing the fundamental ill-posed nature of the inverse problem as well as the epistemic uncertainty induced by limited resources.
One of the uniqueness of the work presented here, is the aspect of the different modules working in conjunction to enable explicit inverse mappings.

\subsection{Forward Modeling}
In the first step of the framework, we train a probabilistic surrogate model of the forward process, using Gaussian process (GP) regression ~\cite{rasmussen2006}. 
Depending on the available sources of the data, we build either a single-fidelity GP or a Multi-Fidelity Gaussian Process (MFGP).

\subsection*{Single-fidelity GP}
Consider a GP surrogate model of the form:
\begin{equation}
y(\bm{x}) \sim  GP(m(\bm{x}),k(\bm{x},\bm{x}^{'} )),
\end{equation}
\noindent where $m(\bm{x})$, assumed to be zero here, is the mean function, and $k(\bm{x},\bm{x}^{’})$ is the covariance function for a vector valued input denoted by $\bm{x}$ of dimension $d$. 
In this work, the covariance function is assumed to be the squared exponential kernel~\cite{rasmussen2006}:

\begin{equation}
k(\bm{x},\bm{x}^{’})=\sigma^2  \exp\left(-\beta (\bm{x}-\bm{x}^{’} )^2 \right) + I \lambda^2,
\label{eq:gp_kernel}
\end{equation}
\noindent where $\beta$ are the (inverse) length scale parameters collected in a vector, one per input dimension, $\sigma^2$ captures the data variance as the amount of data variance captured by the model and $\lambda^2$ quantifies the amount of variance captured by the residuals. 
This GP models an output $y(\bm{x})$ given an input vector $\bm{x}$.
Observe now a set of inputs and outputs and collect these in a training data set of $N$ elements $D=\left\{x_i,y_i \right\}_{i=1}^N$.
 
The GP fitting process translates to fitting the hyperparameters associated with the matrix:
\begin{equation}
K_{(i,j)} = k(x_i,x_j),
\label{eq:gp_cov}
\end{equation}
\noindent where $x_i$ is the $i^{th}$ training datum.

The hyperparameters of the GP are defined as the vector $\bm{\phi} =(\sigma,\beta,\lambda)$ and need to be fitted to the training data set $D$.
In this work, priors are placed on the hyperparameters to incorporate the initial belief into the data modeling before seeing the data itself, such as smoothness.
It can be inferred from above that for our problems of learning the model's hyperparameters, it is always a problem of estimating $m\times (d+2) + 1$ number of parameters. 
The likelihood of observing the training data for a selected set $\btheta$ of hyperparameters is $L(\mathbf{D}|\bphi) = \frac{1}{|\Sigma|^{\frac{1}{2}}}\exp(-\frac{1}{2}\bY^{T}\Sigma^{-1}\bY)$,
where the $i, j$th element of the matrix $\bY$ is the $j$th output value for the $i$th training data point.
The conditional posterior of the hyperparameter set $\bphi$ can then be written as $p(\bphi|\mathbf{D}) \propto L(\mathbf{D}|\bphi)\prod_{k=1}^{m}p(\boldsymbol{\beta}^{(k)})p(\lambda)p(\sigma)$.
This expression, commonly known as the \emph{target distribution}, is known only up to a proportionality constant. 
By combining the priors with the likelihood function, the problem of fitting $\bm{\phi}$ boils down to sampling from the extrema of the posterior distribution $p(\bm{\phi}|D)$ since, of course, larger values of $p(\bm{\phi}|D)$ implies more likely models $\bm{\phi}$. 
The Markov Chain Monte Carlo (MCMC) method both seeks the extrema and provides a way to sample from it, in cases where the normalization constant of the posterior probability distribution is unknown.

\subsection*{Multi-Fidelity Gaussian Process (MFGP)}
In the industrial setting, computational budget allocation might be limited to only a handful of the afforded expensive runs of a high-fidelity simulation code.
On the other hand, access to simplified models (low-fidelity) may provide useful information that at least can capture the general trend of the high-fidelity model. 
The MFGP surrogate can be trained to bridge the information from  various levels of the model's complexity.
To train the MFGP, we follow \highlighttext{Kennedy O’Hagan’s (KOH) methodology~\cite{Kennedy2001}} where the observed data (i.e., the high-fidelity data), $y(x)$, is represented as a linear combination of a low-fidelity and model a discrepancy term~\cite{Kennedy2001}:
\begin{equation}
y(\bm{x})=\eta(\bm{x}; \bm{\theta})+\delta(\bm{x})+ \epsilon,
\label{eq:koh}
\end{equation}
\noindent
where $\bm{\theta}$ are calibration parameters, i.e., parameters of the low-fidelity model that may or may not have a physical meaning, that can be tuned in order to better match the observed data $y(\bm{x})$. 
Note that Eq.~\ref{eq:koh} contains of two separate GPs and a term $\epsilon$ that represents additive Gaussian noise with zero mean and a constant variance. 
Namely, $\eta(\bm{x},\bm{\theta})$ is a GP for the simulator data or the low-fidelity data and $\delta(\bm{x})$  is a GP for capturing the discrepancy between the simulator (or the low-fidelity) and the observed data (or the high-fidelity data) which is collected at the independent variable locations.
Also, we will not consider any calibration parameters ($\bm{\theta}$) with low-fidelity data in the current work, i.e. $\eta(\bm{x},\bm{\theta}) = \eta(\bm{x})$

Importantly, each GP in Eq. \ref{eq:koh} is fitted to its own data set. 
For example, the low-fidelity model $\eta(\bm{x})$ is fitted to a data set $D_{\eta}=\left\{z_i,w_i \right\}_{i=1}^{N_\eta}$ where $z$ is the independent variable and $w$ is the dependent variable (output from the computer simulator). 
The discrepancy $\delta(\bm{x})$ GP is fitted by using information from both the low and the high-fidelity data $D_y=\left\{x_i,y_i \right\}_{i=1}^{N_y}$. 
The KOH method is a two-part solution: build a base model of the low-fidelity data and a discrepancy model that maps the low-fidelity model to high-fidelity data.

Generally, $\bm{z}$ and $\bm{x}$ are not located at the same points, and most typically will not have the same size, i.e., the simulator (low-fidelity model) is run at different input points than the observed (high-fidelity data), but nothing prevents them from being the same.

The covariance matrix on a finite sample of points from both the simulator and the real-world experiment is now given by the following overall structure (the subscript $mf$ refers to multi-fidelity):

\begin{equation}
K_{mf} = \left[
\begin{matrix}
K_y & 0 & 0 \\
0 & K_u & K_{uw} \\
0 & K_{uw}^T & K_w
\end{matrix}
\right],
\end{equation}
\noindent where each covariance matrix has been labeled with a subscript identifying which dependent variable data is being modelled. 
$K_y$ is the covariance matrix of the high-fidelity data, $K_w$ is the covariance of the low-fidelity data and the newly introduced variable $u$ refers to the low-fidelity data predicted on the high-fidelity points and thus $K_{uw}$ is the covariance matrix between the low-fidelity data and the low-fidelity model predicting high-fidelity data. 
This covariance matrix contains multiple hyperparameters via the covariance matrices, which in turn are defined from the covariance function in Eq. \ref{eq:gp_kernel}. 
Similar to single-fidelity GP, the hyperparameters of MFGP are estimated using MCMC, see Ghosh et. al. \cite{ghosh2020advances} for more details, which is a fully Bayesian approach to estimates the posterior distribution of the parameters based on the observed data and prior distribution.

\subsection{Inverse Modeling}
The inverse problem aims to recover the high-dimensional input $\bm{x}\in\mathbb{R}^{M}$ given noisy and gappy data $\tilde{\bm{y}}\in\mathbb{R}^{D}$. This problem is ill-posed since one may not be able to uniquely recover the high-dimensional input $\bm{x}$ given the noisy observations $\tilde{\bm{y}}$. 
To address this challenge, we develop a model that maps the given noisy observations $\tilde{\bm{y}}$ to the unknown high-dimensional input space $\bm{x}$.
In this work, we use a deep generative model (DGM) for constructing the inverse surrogate model.

\subsection*{Conditional Invertible Neural Network (cINN)}

Recently, several conditional deep generative models were developed using generative adversarial networks (GANs)~\cite{goodfellow2014generative} and Variational autoencoders (VAEs)~\cite{kingma2013auto} for solving the computer vision problems~\cite{isola2017image, mirza2014conditional, sohn2015learning}. The main drawbacks of these conditional deep generative models are that it is hard to train the model stably, and it is also difficult to obtain samples with sharp features~\cite{ardizzone2019guided}. Therefore, in this work, we extend the \highlighttext{real-valued non-volume preserving (real NVP)}~\cite{dinh2016density} architecture by a conditioning network that basically conditions the observations $\tilde{\bm{y}}$ to the real NVP~\cite{dinh2016density} architecture. The main advantage of using the real NVP is that it explicitly learns the data distribution with exact log-likelihood evaluation and allow stable training.  
\par
The real NVP architecture transforms a simple distribution $p(\bm{z})$ to a complex distribution $p(\bm{x})$, using a sequence of affine coupling layers. 
Given the latent-variable $\bm{z} \sim p(\bm{z})$, the input space $\bm{x} \sim p(\bm{x})$ can be inferred as $\bm{x}=f(\bm{z})$ and the latent variable can also be calculated inversely with $\bm{z}=f^{-1}(\bm{x})$. 
Here, the transformation function $f$ is invertible, and it is referred to as the affine coupling layer that should satisfy the following properties:  a) the transformation must be invertible and, b) the Jacobian determinant should be easy to compute, and finally, the dimensions of the input and the output must be the same. 
The real NVP consists of forward and inverse propagation. 
During the forward propagation, the model transforms the input $\bm{x}$ to the latent space $\bm{z}$ as follows. 
Given an $M$ dimensional input $\bm{x}$ and $m<M$, the output of an affine coupling layer $\bm{z}$ follows:
\begin{equation}\label{eq: forward_prop}
\begin{array}{l}
\bm{z}_{1: m}=\bm{x}_{1: m}, \\
\bm{z}_{m+1: M}=\bm{x}_{m+1: M} \odot \exp \left(s\left(\bm{x}_{1: m}\right)\right)+t\left(\bm{x}_{1: m}\right).
\end{array}
\end{equation}
During the inverse propagation, the model transforms the latent space $\bm{z}$ to the input space $\bm{x}$ as follows:
\begin{equation}\label{eq: inverse_prop}
\begin{array}{l}
\bm{x}_{1: m}=\bm{z}_{1: m}, \\
\bm{x}_{m+1: M}=\left(\bm{z}_{m+1: M}-t\left(\bm{z}_{1: m}\right)\right) \odot \exp \left(-s\left(\bm{z}_{1: m}\right)\right),
\end{array}
\end{equation}
where the scale $s(\cdot)$ and shift $t(\cdot)$ networks map from $\mathbb{R}^{m}$ to $\mathbb{R}^{M-m}$ and $\odot$ denotes an element-wise multiplication.
The Jacobian of the transformation function $f$ conducted by an affine coupling layer is derived as following~\cite{dinh2016density}:
\begin{equation}
J=\left[\begin{array}{cc}
\mathbb{I}_{m} & \bm{0}_{m \times(M-m)} \\
\frac{\partial \bm{z}_{m+1: M}}{\partial \bm{x}_{1: m}} & \operatorname{diag}\left(\exp \left(s\left(\bm{x}_{1: m}\right)\right)\right)
\end{array}\right],
\end{equation}
Hence the determinant of the Jacobian is simply the product of diagonal elements. 
For complete details on the real NVP architecture, we direct the reader to the work by Dinh et al.~\cite{dinh2016density}.
\par

\subsection*{Network details}
\begin{figure}[H]
	\centering
	\includegraphics[width=0.5\textwidth]{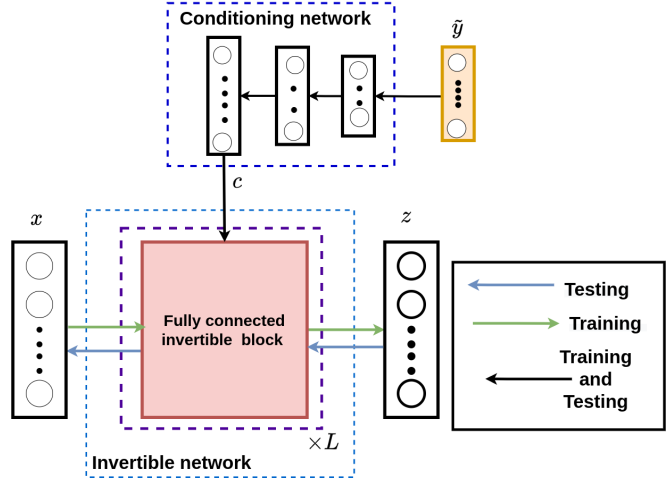}
	\caption{Conditional invertible neural network (cINN): the training process is shown in green color arrow and the testing process is shown in blue arrow.}
	\label{fig:network}
\end{figure}

Both the invertible and the conditioning networks are constructed using fully connected layers. 
Here, the conditioning network takes in the noisy observations $\tilde{\bm{y}}$ as the input and the output being the conditioning inputs $\bm{c}$. 
The invertible network consists of a sequence of affine coupling layers with $L$ blocks that transforms the input space $\bm{x}$ to the latent space, $\bm{z}$ with the conditioning inputs $\bm{c}$ from the conditioning network during forward propagation. 
Here, during forward propagation, we denote  $\{\bm{x}^{\prime}_{l-1}\}_{l=1}^{L} \in \mathbb{R}^{M}$ as the input to each invertible blocks and the output for that invertible block is $\{\bm{x}^{\prime}_l\}_{l=1}^{L} \in \mathbb{R}^{M}$. During the inverse propagation, the model transforms the latent space $\bm{z}$ to the input space $\bm{x}$ conditioned on the conditioning inputs $\bm{c}$. Here, we denote $\{\bm{z}^{\prime}_{l-1}\}_{l=1}^{L}$ as the input and the output as $\{\bm{z}^{\prime}_{l}\}_{l=1}^{L}$ for that invertible block during inverse propagation. 
Figure ~\ref{fig:network} illustrates our inverse surrogate model showing the training (forward propagation) and the testing (inverse propagation) process.

\par

\begin{figure*}[]
    \centering
    \includegraphics[width=0.7\textwidth]{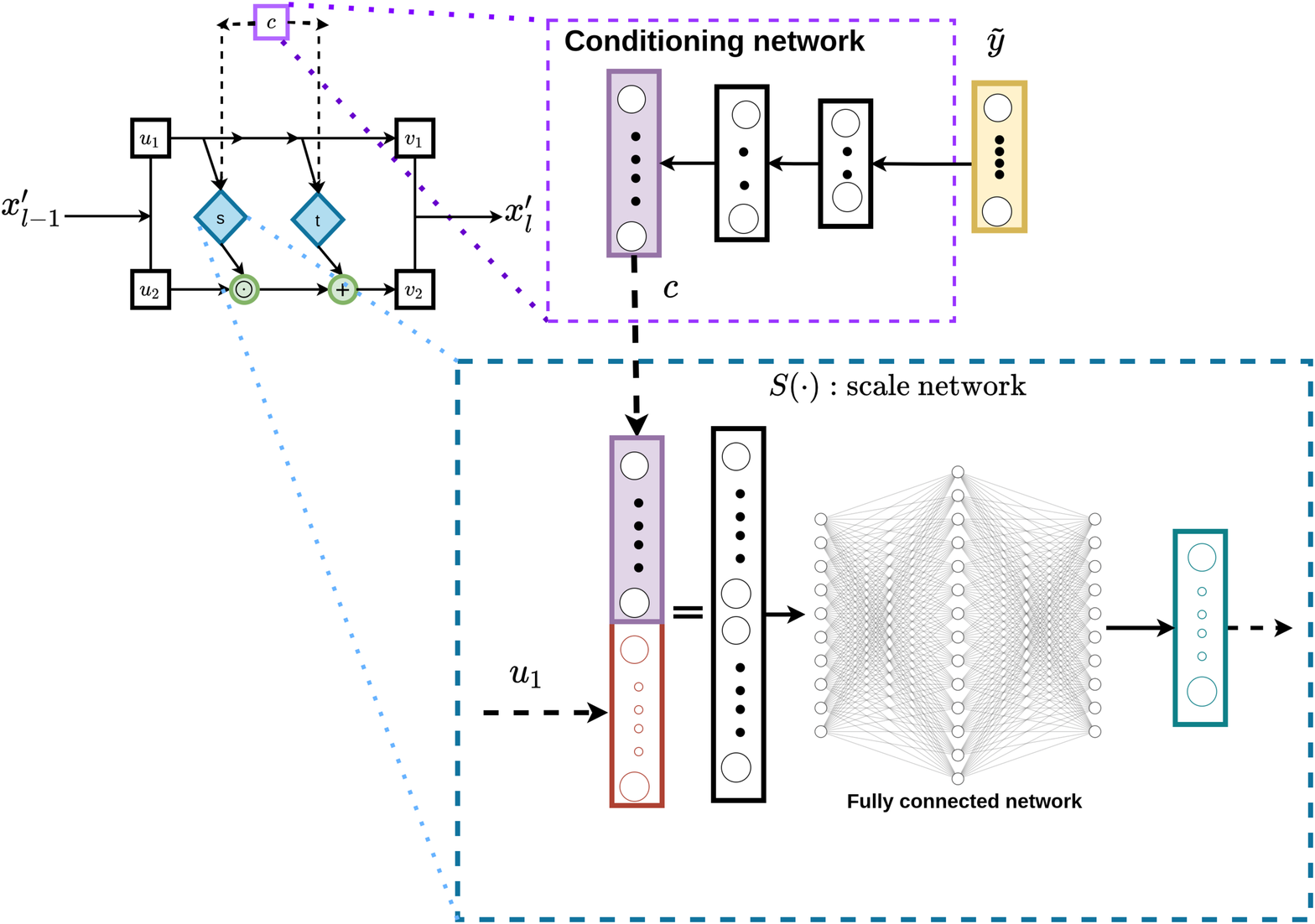}
    \caption{Illustration of a scale network, $s(\cdot)$ in an invertible block.  Here, $\bm{u}_1 = \bm{x}^{\prime}_{{(1: m)}, \{l-1\}}$, $\bm{u}_2 = \bm{x}^{\prime}_{{(m+1: M)}, \{l-1\}}$, $\bm{v}_1 = \bm{x}^{\prime}_{{(1: m)}, \{l\}}$ and $\bm{v}_2 = \bm{x}^{\prime}_{{(m+1: M)}, \{l\}}$}
    \label{fig:network_affine}
\end{figure*}

\highlighttext{The concept of concatenating the conditioning inputs to the invertible network blocks are similar to the work of Padmanabha and Zabaras~\cite{padmanabha2020solving}, Geneva and  Zabaras~\cite{geneva2020multi}, Zhu et al.~\cite{zhu2019physics} and Ardizzone et. al.~\cite{ardizzone2019guided}. However, we modify the conditioning and the invertible architectures to construct an inverse surrogate model that maps the noisy observations to the input space. 
The details of forward and inverse propagation of each affine coupling block can be found in Table \ref{tab:affine_layer}.} \highlighttext{Figure~\ref{fig:network_affine} illustrates the single affine conditional block where the conditioning inputs $\bm{c}$ from the conditioning network and the input $\bm{x}_{l-1}$ for the invertible block $l$ are considered as the input to the scale network $s(\cdot)$. Similarly, we consider the same input for shift $t(\cdot)$ network. In this work, both the scale and the shift networks are fully connected layers with LeakyReLU~\cite{maas2013rectifier} activation function between each hidden layer.} \highlighttext{In this work, we empirically perform an extensive search for network architecture designs such as the number of affine coupling layers, and the numbers of hidden layers, and an extensive hyperparameters search such as the learning rate, weight decay, and the number of epochs that work well for this inverse problem.}

\highlighttext{To clarify, the invertible network or the block of the cINN architecture is the sequence of affine coupling layers (one shown in Figure ~\ref{fig:network}) which maps the input space $\bm{x}$ to latent space $\bm{z}$.
To account for the observation $\bm{y}$ during training (and target objective during the inverse process), a conditioning network is used which maps $\bm{y}$ to a conditioning input $\bm{c}$.  The conditioning input $\bm{c}$ becomes input to the  scale $s(\cdot)$ and the shift $t(\cdot)$ network as shown in Figure~\ref{fig:network_affine}.}

\begin{table}[H]
	\caption{Forward and inverse propagation of each affine coupling layer conditioned on the conditioning input $\bm{c}$. During forward propagation, the input is $\{\bm{x}^{\prime}_{l-1}\}_{l=1}^{L} \in \mathbb{R}^{M}$ and the output is $\{\bm{x}^{\prime}_l\}_{l=1}^{L} \in \mathbb{R}^{M}$. During inverse propagation, the input is the latent space $\{\bm{z}^{\prime}_{l-1}\}_{l=1}^{L}$ and the output is $\{\bm{z}^{\prime}_{l}\}_{l=1}^{L}$. Here, the split operation divides the data into two parts: $1<m<M$, such that $\bar{\bm{x}}_{1,{l-1}}$ = $\bm{x}^{\prime}_{(1:m),\{{l-1}\}}$ and $\bar{\bm{x}}_{2,{l-1}}$ = $\bm{x}^{\prime}_{(m+1:M),\{{l-1}\}}$ during the forward operation. Similarly, $\bar{\bm{z}}_{1,{l-1}}$ = $\bm{z}^{\prime}_{(1:m),\{{l-1}\}}$ and $\bar{\bm{z}}_{2,{l-1}}$ = $\bm{z}^{\prime}_{(m+1:M),\{{l-1}\}}$ for the inverse operation. \highlighttext{Here, AffineCouplingNet represents the scale $s(\cdot)$ and shift $t(\cdot)$ networks.}} 
	\centering
	\begin{tabular}{l|l}
		Forward & Inverse \\ \hline
		$\overline{\bm{x}}_{1,{l-1}}, \overline{\bm{x}}_{2,{l-1}}$={split}$\left(\bm{x}^{\prime}_{{l-1}}\right)$&  $\overline{\bm{z}}_{1,{l-1}}, \overline{\bm{z}}_{2,{l-1}}$={split}$\left(\bm{z}^{\prime}_{l-1}\right)$\\
		$\hat{\bm{x}}_{1,{l-1}}$={concat}$\left(\overline{\bm{x}}_{1,{l-1}}, \bm{c}\right)$ &  $\hat{\bm{z}}_{1,{l-1}}$={concat}$\left(\overline{\bm{z}}_{1,{l-1}}, \bm{c}\right)$\\
		\small $({\bm{s}}, \bm{t})$={AffineCouplingNet}$\left(\hat{\bm{x}}_{1,{l-1}}\right)$&  \small$({\bm{s}}, \bm{t})$={AffineCouplingNet}$\left(\hat{\bm{z}}_{1,{l-1}}\right)$ \\
		$\overline{\bm{x}}_{2,{l}}=exp(\bm{s}) \odot \overline{\bm{x}}_{2,{l-1}}+\bm{t}$ & $\overline{\bm{z}}_{2,l}=\left(\overline{\bm{z}}_{2,{l-1}}-\bm{t}\right) / exp(\bm{s})$ \\
		$\overline{\bm{x}}_{1,{l}}=\overline{\bm{x}}_{1,{l-1}}$& $\overline{\bm{z}}_{1,{l}}=\overline{\bm{z}}_{1,{l-1}}$  \\
		$\bm{x}^{\prime}_{l}$={concat}$\left(\overline{\bm{x}}_{1,{l}}, \overline{\bm{x}}_{2,{l}}\right)$& $\bm{z}^{\prime}_{l}$={concat}$\left(\overline{\bm{z}}_{1,{l}}, \overline{\bm{z}}_{2,{l}}\right)$
	\end{tabular}
	\label{tab:affine_layer}
\end{table}

\subsubsection*{Loss function:}
Given a set of training data  $\mathcal{D} = \{\bm{x}^i,\bm{\tilde{y}}^i\}_{i=1}^{N}$, we consider the \textit{maximum a posteriori} (MAP) estimate of the model parameters $\bm{\theta}$:
\begin{equation}\label{eq:MAP_1}
\bm{\theta}^{*} = \arg \max _{\bm{\theta}} \prod_{i=1}^{N}  p(\bm{\theta} | \bm{x}^{(i)}, \tilde{\bm{y}}^{(i)}).
\end{equation}
In this work, the invertible and the conditional networks are trained in an end to end fashion. Therefore, the model parameters, $\bm{\theta}$, include the invertible network parameters $\bm{\theta}_I$ and the conditioning network parameters $\bm{\theta}_c$, i.e., $\bm{\theta} = [\bm{\theta}_c, \bm{\theta}_I]$. 
We further simplify Eq.~\ref{eq:MAP_1} as follows:
\begin{equation}\label{eq:MAP_2}
\bm{\theta}^{*} = \arg \max _{\bm{\theta}} p(\bm{\theta}) \prod_{i=1}^{N} p(\bm{x}^{(i)}|\tilde{\bm{y}}^{(i)}, \bm{\theta}),
\end{equation}
where $p(\bm{\theta})$ is the prior on the model parameters and  the conditional likelihood is simplified as follows. 
First, we introduce a bijection ${f}_{\bm{\theta}}(\cdot)$ that is parametrized by both the invertible and the conditioning model parameters $\bm{\theta} = [\bm{\theta}_c, \bm{\theta}_I]$ and conditioned on the noisy observations $\tilde{\bm{y}}$, that maps  the input space $\bm{x}$ to the latent space $\bm{z}\sim p(\bm{z})$. 
Next, using the change of variables formula, we obtain the following conditional likelihood:
\begin{equation}
p_{\bm{\theta}}(\bm{X} \mid \tilde{\bm{Y}})= \prod_{i=1}^{N} \left[ p(f_{\bm{\theta}}(\bm{x}^{(i)},\tilde{\bm{y}}^{(i)})) \cdot \left|\operatorname{det}\left(\frac{\partial( f_{\bm{\theta}}(\bm{x}^{(i)},\tilde{\bm{y}}^{(i)}))}{\partial \bm{x}^{(i)}}\right)\right| \right],
\end{equation}
where, $\bm{X}=\left\{\bm{x}^{(1)}, \bm{x}^{(2)}, \ldots, \bm{x}^{(N)}\right\}$, $\tilde{\bm{Y}}=\left\{\tilde{\bm{y}}^{(1)}, \tilde{\bm{y}}^{(2)}, \ldots, \tilde{\bm{y}}^{(N)}\right\}$ and the latent variable $\bm{z}^{(i)} = f_{\bm{\theta}}(\bm{x}^{(i)},\tilde{\bm{y}}^{(i)})$. 

The conditional log-likelihood $\log p_{\bm{\theta}}({\bm{X}} \mid \tilde{\bm{Y}})$ can be exactly evaluated as follows:
\begin{equation} \label{cond_LL}
\log p_{\bm{\theta}}(\bm{X} \mid \tilde{\bm{Y}})= \sum_{i=1}^{N} \log p(f_{\bm{\theta}}(\bm{x}^{(i)},\tilde{\bm{y}}^{(i)})) + \log \left|\operatorname{det}\left(\frac{\partial( f_{\bm{\theta}}(\bm{x}^{(i)},\tilde{\bm{y}}^{(i)}))}{\partial \bm{x}^{(i)}}\right)\right|.
\end{equation}   
One can re-write the MAP estimate in Eq.~\ref{eq:MAP_2} can be written as the minimizing the below loss function:
\begin{equation}\label{eq:loss3}
\begin{split}
\mathcal{L}= - \sum_{i=1}^{N} \left[\log p(f_{\bm{\theta}}(\bm{x}^{(i)},\tilde{\bm{y}}^{(i)})) + \log \left|\operatorname{det}\left(\frac{\partial( f_{\bm{\theta}}(\bm{x}^{(i)},\tilde{\bm{y}}^{(i)}))}{\partial \bm{x}^{(i)}}\right)\right|\right]\\
- \log p(\bm{\theta}).
\end{split}
\end{equation}   
Since we consider a standard normal distribution for the latent space, we can further simplify the $\log p(f_{\bm{\theta}}(\bm{x}^{(i)},\tilde{\bm{y}}^{(i)}))$ as $-\frac{\left\|f_{\bm{\theta}}(\bm{x}^{(i)},\tilde{\bm{y}}^{(i)})\right\|_{2}^{2}}{2}$. 
\\
As mentioned above, we consider a Gaussian prior on the model parameter with mean $0$ and variance $\sigma^{2}_{\bm{\theta}}$. 
Further, simplifying the Eq.~\ref{eq:loss3}, we obtain the following loss function:
\begin{equation} \label{eq:loss_eq}
\mathcal{L}=\frac{1}{N}\sum_{i=1}^{N}\left[\frac{\left\|f_{\bm{\theta}}(\bm{x}^{(i)},\tilde{\bm{y}}^{(i)})\right\|_{2}^{2}}{2}-\log \left|J_{i}\right|\right]+\tau\|\bm{\theta}\|_{2}^{2},
\end{equation}
where, $\tau = 1/2\sigma^2_{\bm{\theta}}$ and the Jacobian determinant $J_{i} = \operatorname{det}\left(\frac{\partial( f_{\bm{\theta}}(\bm{x}^{(i)},\tilde{\bm{y}}^{(i)}))}{\partial \bm{x}^{(i)}}\right)$. The training process is enumerated in Algorithm~\ref{algo_train}.
\begin{algorithm}[h]
	\caption{Training procedure for the inverse surrogate model.}
	\label{algo_train}
	\KwIn{Training data: $\{\bm{x}^{(i)}, \tilde{\bm{y}}^{(i)}\}_{i=1}^{N}$,  number of epochs: $E_{\text{train}}$, learning rate: $\eta$, mini-batch size: $Q$, and number of flow model: $L$.}
	\For{epoch =  $1$ to $E_{\text{train}}$}{
		Sample a minibatch from the training datset: $\{\bm{x}^{(i)}, \tilde{\bm{y}}^{(i)}\}_{i=1}^{Q}$ and pass the observations $\{\tilde{\bm{y}}^{(i)}\}_{i=1}^{Q}$ to the conditioning network to obtain the conditioning inputs $\{{\bm{c}}^{(i)}\}_{i=1}^{Q}: \bm{c}^{(i)} = g_{\bm{\theta}_c}(\tilde{\bm{y}}^{(i)})$ and the input $\{\bm{x}^{(i)}\}_{i=1}^{Q}$ to the conditional invertible network: \\
		\For{$l$ = $1$ to $L$}{
			$\bm{x}^{\prime(i)}_{0} = \bm{x}^{(i)}$ \\
			{$\bm{x}^{\prime(i)}_l, {\text{J}}^{\prime(i)}_l = {f}_{l,\bm{\theta}_I}(\bm{x}^{\prime(i)}_{(l-1)},{\bm{c}}^{(i)})$} \Comment{${f}$ is the invertible network that includes permutation.}} 
		$\hat{\bm{z}}^{(i)}$ = $\{\bm{x}^{\prime(i)}_{L}\}$ \\
		Compute $\text{J}_{\text{final}}^{(i)} = \sum_{l=1}^{L}({\text{J}}^{\prime(i)}_{l})$ \\
		$\mathcal{L}$ = Loss($\hat{\bm{z}}^{(i)},\text{J}_{\text{final}}^{(i)}$) \\
		$\nabla \bm{\theta}\leftarrow$Backprop$\left(\mathcal{L}\right)$ \\
		$\bm{\theta} \leftarrow \bm{\theta}-\eta \nabla \bm{\theta}$
	}
	\KwOut{Trained cINN network.}
\end{algorithm}

\subsubsection*{Inversion:}
After the training process, we obtain the inverse solutions to the cINN as illustrated in Algorithm~\ref{algo_test}. 
Given the observation $\tilde{\bm{y}}$ as the input to the conditioning network, we get the conditioning input $\bm{c}$ to the invertible network. 
For the invertible network, we first generate $S=1000$ samples: $\mathcal{\bm{Z}} = \{\bm{z}^{(j)}\}_{j=1}^{S}$, where $\bm{z}$ follows normal distribution as $\bm{z}^{(j)}\sim \mathcal{N}(0,1)$ and then we obtain the input field $\{\hat{\bm{x}}^{(j)}\}_{j=1}^{S}$ using invertible network conditioned on the observations $\tilde{\bm{y}}$:
\begin{equation}\label{inverse_map}
\hat{\bm{x}}^{(j)} = f_{\bm{\theta}}^{-1} (\bm{z}^{(j)},\tilde{\bm{y}}).
\end{equation}
\begin{algorithm}[h]
	\caption{Inverse solution: Conditional invertible neural networks.}
	\label{algo_test}
	\KwIn{Trained invertible and conditioning networks, observations: $\tilde{\bm{y}}$, number of samples: $S$, and number of flow model: $L$.
		\SetAlgoLined \\
		$\bm{z}^{(j)}\sim \mathcal{N}(0,1)$; $\mathcal{\bm{Z}} = \{\bm{z}^{(j)}\}_{j=1}^{S}$. \\
		${{\bar{\bm{y}}}} = \text{tile}(\tilde{\bm{y}})$; ${\bm{\bar{y}}} = {\{\tilde{\bm{y}}^{(j)}\}_{j=1}^{S}}$   \Comment{tile: constructs a new array by repeating $\tilde{\bm{y}}$} \\  
		Pass the observations $\{\tilde{\bm{y}}^{(i)}\}_{j=1}^{S}$ to the conditioning network to obtain the conditioning inputs $\{{\bm{c}}^{(i)}\}_{j=1}^{S}: \bm{c}^{(i)} = g_{\bm{\theta}_c}(\tilde{\bm{y}}^{(i)})$
		and the samples ${\{\bm{z}^{(j)}\}}_{j=1}^{S}$ to the invertible network:}
	\For{$l$ = $1$ to $L$}{
		$\bm{z}^{\prime(j)}_0 = \bm{z}^{(j)}$ \\
		$\bm{z}^{\prime(j)}_l = {f}^{-1}_{l,\bm{\theta}_I}(\bm{z}^{\prime(j)}_{(l-1)},{\bm{c}}^{(j)})$ \Comment{${f}^{-1}$ is the inverse mapping that includes permutation}} 
	\hspace{0.25cm}$\hat{\bm{x}}^{(j)} = \bm{z}^{\prime(j)}_L$ \\
	{\textbf{Output:} Samples:$\{\hat{\bm{x}}^{(j)}\}_{j=1}^{S}$}
\end{algorithm}

\section{DEMONSTRATION EXAMPLE: TOY FUNCTION}

We now consider the application of Algorithms \ref{algo_train} and \ref{algo_test} on a simple low-dimensional case.
We consider an input domain 
$\Omega_x = [-L_x/2, L_x/2]^{d_x}$
equipped with a uniform probability density $p(x)$ over $\Omega_x$.
The ground truth forward function is modeled as $y=f(\vc x) + \epsilon$, where $f(\cdot)$ is a simple quadratic, 
\begin{equation}
	f(\vc x)=(\mtx W \vc x - \vc \mu)^\intercal (\mtx W \vc x - \vc \mu),
	\label{eqn:toy_forward}
\end{equation}
and $\epsilon \sim \scriptN(0,0.5^2)$.
Equation\ (\ref{eqn:toy_forward}) corresponds to a second-order approximation of the neighborhood of any local extremum in a twice-differentiable function via appropriate choice of the affine transformation parameters $\mtx W$ and $\vc \mu$.
This makes it relevant to many engineering design and optimization tasks, where one is interested in locating such an extremum in design space (maximizing efficiency, minimizing weight, etc).
For our example, we choose $d_x=2$ input dimensions, $L_x=4$, $\mtx W = \mtx I_{d_x \times d_x}$, and $\vc \mu = \vc 0_{d_x}$.
This means that the unique global minimum of $f(\vc x)$ is $\vc x = \vc 0$ with value $y=0$.
Furthermore, the set of inputs consistent with any $y\ge 0$ is the circle of radius $y^{1/2}$ centered at the origin.

\begin{figure}[hbt]
	\begin{center}
		\includegraphics[width=0.5\textwidth]{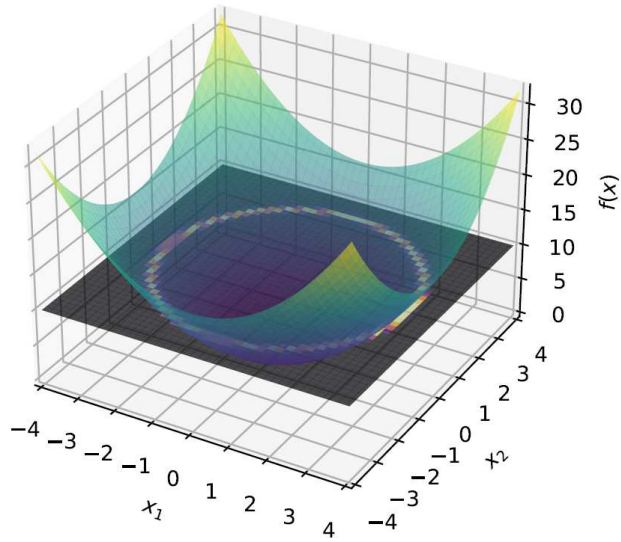}
	\end{center}
	\caption{INN toy example: Plot of the two-dimensional function along with a level-set at $y=10$.}
	\label{fig:toy_surface}
\end{figure}

An illustration of $f(\cdot)$ with a slice at $y=10$ showing the ground truth inverse probability $p(\vc x | y)$ convolved with a mild Gaussian smoother (for visualization purposes) is shown in Figure\ \ref{fig:toy_surface}.
We see that, physically, the \textit{true} conditional distribution that we wish to capture with the cINN is ring-shaped.
Intuitively, \textit{any} $\vc x$ on this ring results in the specified $y$, reflecting the fundamental one-to-many nature associated with the inverse problem in general.
In other words, the marginal densities $p(x_i|y)$ will \textit{not} necessarily converge to having zero standard deviation with sufficiently many data.

We train a cINN with $L=4$ affine coupling pairs using Algorithm \ref{algo_train} with minibatches of $Q=128$ data produced in an online setting by sampling $\vc x \sim p(x)$, $y = f(\vc x)+\epsilon$.
A flat weighting function is used.
We optimize for $2 \times 10^4$ iterations using the Adam optimizer \cite{kingma2014adam} with cosine-annealed learning rate from $\eta=3\times10^{-3}$ to $\eta=10^{-5}$ by the last iteration.

\begin{figure}[hbt]
	\begin{center}
		\includegraphics[width=0.45\textwidth]{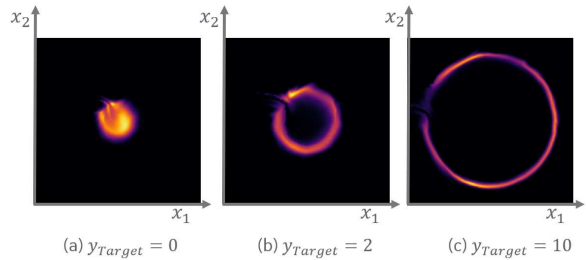}
	\end{center}
	\caption{INN toy example: Conditional distributions learned by the model for (left to right) $y=0$, $2$, and $10$.}
	\label{fig:toy_samples}
\end{figure}

\highlighttext{With the trained model, a designer can query the cINN  with different target objectives ($y$) to generate designs solutions $\vc x$. 
Heat maps of the conditional distribution $p(\vc x | y)$ learned by the cINN are plotted in Figure\ \ref{fig:toy_samples} for various values of $y$.
All the samples from the distributions are possible design for corresponding design target.  
We see that the cINN has successfully learned to model the expected conditional distribution, in particular one-to-many solution of the toy problem.
Because the cINN explicitly models these conditional distributions, we can trivially solve any design problem, i.e.\ find any number of designs $\vc x$ that are consistent with an arbitrary desired output $y$. By contrast, a traditional forward modeling-based approach would require many repeated queries to implicitly determine even a single $\vc x$ such that $f(\vc x)=y$.
From the designer's perspective, one may have multiple design choices which satisfies the performance requirement. One can then chose the final design based on additonal preference, desirability towards other requirements or constraints.
}

\section{INVERSE AERODYNAMIC DESIGN OF 3D TURBINE BLADES}

In this section, we will present the process of inverse aerodynamic design of 3D turbine last stage blade.

\subsection*{Modeling and Simulations}
\begin{figure}[h]
	\centering
	\includegraphics[scale=0.4]{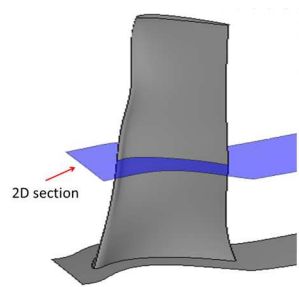}
	\caption{Construction of 3D airfoil based on 2D sections.}
	\label{fig:3Dairfoil_parametrization}
\end{figure}

\begin{figure}[t]
	\begin{center}
		\setlength{\unitlength}{0.012500in}
		\includegraphics[width=0.55\textwidth]{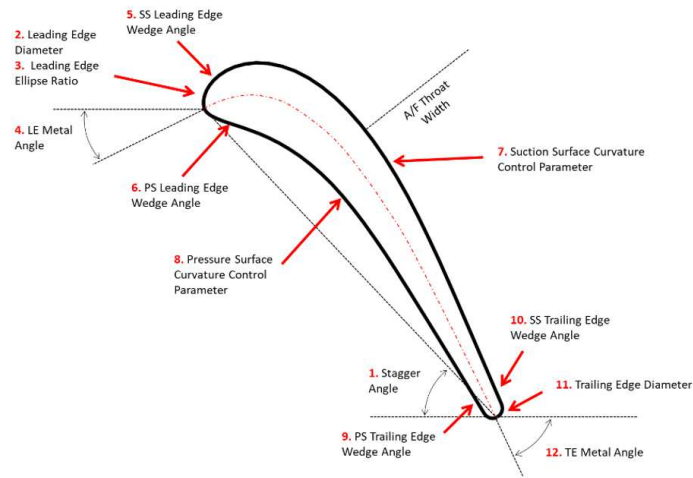}
	\end{center}
	\caption{ 2D airfoil parametrization.}
	\label{fig:2Dpar} 
\end{figure}

The reference design and the turbine operating conditions for modeling and simulation of 3D blade are based on an industrial gas turbine last stage.
Turbine blade performance is evaluated using steady-state RANS CFD at two different fidelity levels; a fast running coarse mesh for broader design space exploration, and a slower running fine mesh for accuracy refinement. 
The 3D blade surface is constructed with seven 2D airfoil profiles at different spanwise locations from hub to tip. 
Figure \ref{fig:3Dairfoil_parametrization} shows a view of the full 3D blade and a 2D sectional plane represented by an airfoil.
Each airfoil section is characterized by 12 independent parameters, as previously shown in Figure \ref{fig:2Dpar}.
These parameters allow independent control of the stagger angle, leading and trailing edge metal angles, leading edge diameter, suction and pressure side wedge angles, leading edge and trailing edge metal angles, and the airfoil curvature between the leading and trailing edges.
The ranges for these parameters are selected to provide a wide design space while respecting geometrical constraints.

The 2D sections are aligned relative to each other in circumferential and axial space by aligning the section centers of gravity (CG) along a radial line through the hub section CG (referred to as the stacking line). After the section CGs are aligned, one additional parameter, referred to as the airfoil lean angle, is applied to reorient the stacking line relative to the radial direction. Surfaces are fit through the seven stacked sections to create the full, continuous, 3D airfoil definition. Eighty-five total parameters are therefore required to define the complete 3D airfoil shape.

The parameters are expressed as offsets from a baseline, requiring that each section starts from an appropriate reference design. An in-house software package tailored specifically for turbomachinery design uses the parameters described above to create the airfoil coordinates, and these coordinates are then transformed into a full 3D CAD model of the rotor blade for the CFD grid generation. A 3D structured mesh is built using a commercially available software package. Grid templates are built from the baseline geometry and used consistently for all the cases throughout the optimization. With this approach, all the grids have similar refinement and quality metrics. To simulate the full stage, the upstream stator is included in the CFD calculation for each case. The design of the vane and the vane mesh are not altered through the optimization.

\begin{figure}[h]
	\centering
	\includegraphics[width=0.4\textwidth]{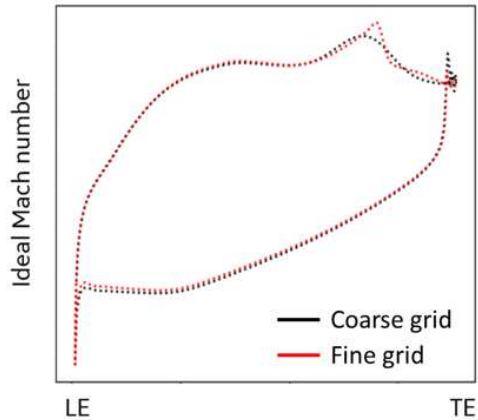}
	\caption{Comparison between coarse and fine grid resolutions for the baseline case.}
	\label{fig:3D_CFD_GridSensitivity}
\end{figure}

The 3D CFD analysis is performed using GE’s in-house CFD solver TACOMA, a 2nd-order accurate (in time and space), finite-volume, block-structured, compressible flow solver. The steady RANS calculations are solved with a mixing plane between rotating and stationary components. Source terms are included at various locations along the end walls to simulate the injected cooling, leakage, and purge flows.
The multi-fidelity dataset is based on two sets of grids; a low resolution (low node count) grid and a high resolution grid, to obtain low and high fidelity data respectively. Several grids are assessed to down select coarse and fine grid resolutions that provide a sufficiently wide trade off between fidelity and computational cost. A production-level grid is selected to provide accurate performance metrics for the high-cost data.
The grid is then progressively relaxed to reduce computational cost for the low-cost cases.
Figure \ref{fig:3D_CFD_GridSensitivity} shows the ideal Mach number at the tip for the baseline geometry, assessed using the coarse and the fine grid, i.e. low and high fidelity respectively. The ideal Mach number distributions agree very well at all locations, except for the rear portion of the suction side. Due to the poorer resolution of the coarse gird, the shock is not accurately captured in the low fidelity calculations.
Considering all operations involved in the assessment of each case, starting from the selection of the parameters to the calculation of the objectives, the cost ratio between high and low fidelity simulations is 4.5 to 6.0. 
Some operations, like geometry generation and post-processing, can be considered approximately grid independent.

Two main objectives are selected as metrics of the 3D blade's performance. These scalar targets are calculated from the CFD results, and will be used as inputs in the inverse design approach:
\begin{itemize}
	\item - Scalar Objective 1: Aerodynamic efficiency. The efficiency is calculated as the ratio of mechanical and ideal power. All the inputs required to calculate the efficiency value are directly available as output quantities of the CFD simulation. Generally, in a design process, the preference is to maximize efficiency.
	
	\item -  Scalar Objective 2: Pseudo-reaction or degree of reaction. In a stage calculation, the degree of reaction indicates the split of flow acceleration between stator and rotor. The pseudo-reaction is calculated for each geometry and is monitored to account for the changes in turbine operating condition due to the blade shape. The design preference is normally to target a baseline value.
\end{itemize}

\begin{figure}[h]
	\centering
	\includegraphics[width=0.5\textwidth]{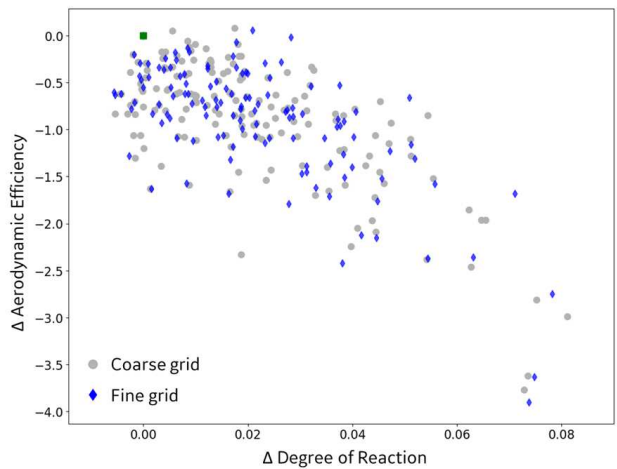}
	\caption{DOE results for coarse and fine grid cases.}
	\label{fig:3D_CFD_DOE}
\end{figure}

The two main objectives, efficiency and degree of reaction, are plotted in  Figure \ref{fig:3D_CFD_DOE}, using the date from initial rounds of DOE.
The cases assessed with a low fidelity grid are represented with gray circles, while the cases assessed with high fidelity are represented with blue diamonds.
The green square represents the performance of the baseline design.
\highlighttext{Figure \ref{fig:3D_CFD_DOE} shows that the range of outputs generated by low-fidelity are very similar to high-fidelity analysis and are correlated. Although, low-fidelity analysis is not very accurate, it stores information which can be used by MFGP model to build accurate model by trading off cost with accuracy. }

\begin{figure}[h]
	\centering
	\includegraphics[width=0.55\textwidth]{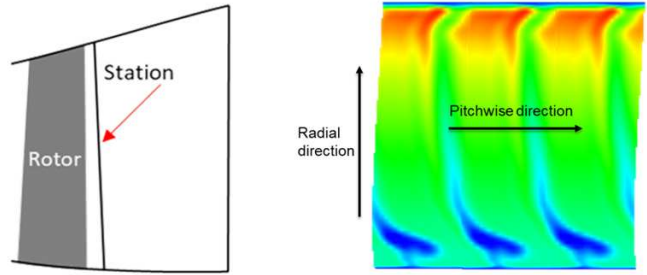}
	\caption{Profile of absolute flow angle downstream of the rotor}
	\label{fig:CFD_profiles}
\end{figure}

Additionally, radial profiles of flow quantities, like pressure and flow angle, are obtained to characterize the quality of flow field propagating from the turbine to the downstream exhaust diffuser, which has not been modeled here.
Additional objectives are formulated based on the desirability of the turbine exit flow profile. 
The focus is on the profiles of absolute total pressure and flow tangential angle, which are expected to have the most important effects on the diffuser performance. These flow quantities affect the momentum distribution going into the diffuser and the flow incidence to the downstream struts.
Figure \ref{fig:CFD_profiles} illustrates the location where the profiles are extracted, downstream of the trailing edge. The right plot shows the rotor wake (repeated by applying periodicity), and shows the direction of the averaging. At each radial location, the quantities are averaged in the pitchwise direction over a single passage, so that the radial profiles include the effect of wakes and shock waves. The flow profiles are extracted downstream of the rotor trailing edge to assess the potential effect on the performance of the downstream exhaust diffuser.

A preliminary analysis of the CFD database showed that the rotor airfoil shape can impose consequential changes to the swirl profile, both in terms of mean value and variability, requiring characterization and objective control in the cINN.

To produce a turbine design which can couple with the diffuser with a satisfactory performance, we want to be able to control the behavior of pressure and swirl close to the end wall, in the boundary layer region, and at midspan. 
To understand what features drive diffuser performance, we leverage the diffuser CFD models and apply a subset of  the DOE turbine exit flow profiles as diffuser inlet conditions. 
Representative merit criteria were extracted for the rotor profiles and correlated to the diffuser performance in terms of recovery factor. 
This approach is useful if we want to characterize the rotor exit profile in terms of scalar metrics.  Additionally, the framework has the capability to impose the profiles as vectors, point by point. 
This capability can be used if a precise inlet flow field is required at the inlet of the diffuser, and the description through scalar metrics is not sufficiently precise.

\subsection*{Forward Modeling}
As discussed in the previous section, the design space of forward process is 85 dimensional. 
The output consists of both scalar and vector variables.
The scalar variables consist of efficiency and pseudo-reaction.
The vector outputs are exit flow profiles given by pressure ($\mathbb{R}^{100 \times 1}$) and swirl angle profiles ($\mathbb{R}^{ 100 \times 1}$), which are pressure and swirl angle measurements at 100 discretized span-wise location.
For the vector outputs, dimensionality reduction is carried out using Principal Component Analysis (PCA), to reduce the dimension of encoded profiles ($\mathbb{R}^{200 \times 1}$) to the lower dimension ($\mathbb{R}^{ 6\times 1}$) represented by the PCA coefficients, capturing $>90\%$ energy of original variables.
Next, all the scalar objectives and PCA coefficients of the vectors are used to build MFGP model using GEBHM.

\highlighttext{In this work, we used multi-fidelity  adaptive sampling \cite{ghosh2019strategy} to pick designs and fidelity of CFD simulations to create the DOE for MFGP training. The new designs of the DOE are adaptively picked based on the cost ratio as well as the amount of uncertainty reduction associated with high and low fidelity simulations. With the goal of achieving less than $10\%$ of nRMSE, a total of 249 low-fidelity and 231 high-fidelity CFD simulation were required. 
The parameters of MFGP are estimated using MCMC \cite{ghosh2020advances}, which is a fully Bayesian approach to estimates the posterior distribution of the parameters based on the observed data and prior distribution. Therefore, it quantifies the uncertainty on the MFGP parameters associated with the sparse data and performs robustly in terms of predictive capability.}

We estimated the accuracy with Coefficient of determination ($R^{2}$) and  Normalized Root Mean Square Error (nRMSE). 
nRMSE is Root Mean Square Error (RMSE) normalized by the difference between maximum and minimum values of output, i.e. $nRMSE = RMSE/ (y_{max} - y_{min})$.
To test the accuracy of the model,  around 10\% of high-fidelity data were hold-out to measure the R-squared and nRMSE. 
The overall accuracy in terms of R-squared and nRMSE of all the models of forward process is given in Table \ref{tab:det_accuracy_mfgp}.

\begin{table}[]
	\caption{Accuracy Metrics: Normalized root mean squared error (nRMSE) and $R^{2}$ score for the aerodynamic efficiency, the pseudo-reaction, and the $6$ PCA coefficients (for the pressure and the swirl angle profiles)} 
	\centering
	\begin{tabular}{c|c|c|}
		\cline{2-3}
		& \textbf{$R^{2}$} & \textbf{nRMSE } \\ \hline \hline
		\multicolumn{1}{ |c|}{Efficiency} & 0.792   & 0.094  \\ \hline
		\multicolumn{1}{|c|}{Pseudo reaction}   & 0.909   & 0.083  \\ \hline
		\multicolumn{1}{|c|}{PCA-1}      & 0.965   & 0.063 \\ \hline
		\multicolumn{1}{|c|}{PCA-2}      & 0.920   & 0.069 \\ \hline
		\multicolumn{1}{|c|}{PCA-3}      & 0.850   & 0.103  \\ \hline
		\multicolumn{1}{|c|}{PCA-4}      & 0.781   & 0.109 \\ \hline
		\multicolumn{1}{|c|}{PCA-5}      & 0.357   & 0.199 \\ \hline
		\multicolumn{1}{|c|}{PCA-6}      & 0.389   & 0.169  \\  \hline
	\end{tabular}
	\label{tab:det_accuracy_mfgp}
\vspace{-0.8cm}
\end{table}

\begin{figure}[h]
	\centering
	\includegraphics[width=0.5\textwidth]{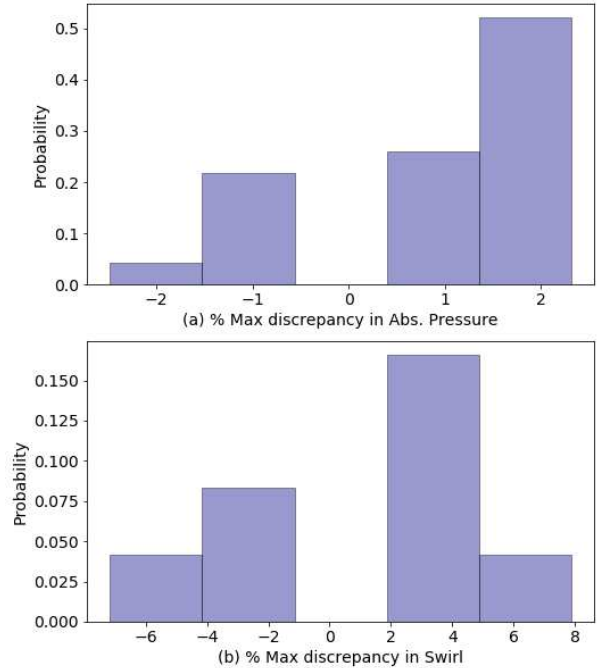}
	\caption{Maximum percentage deviation of profiles predicted by combination of PCA models of vector objective when compared to true values of validation data set of high-fidelity CFD simulations}
	\label{Fig:profile_error}
\end{figure}

\highlighttext{For the scalar objectives, (efficiency and pseudo-reaction), nRMSE of less than $10\%$ was achieved, which has been the requirement from design standpoint. 
The model of the first four PCA coefficients achieved nRMSe of $10\%$ or better. 
It must be noted that the first four PCA coefficients captures more than $80\%$ variability of the vector profiles. 
The last two PCA coefficients (PCA-5 and PCA-6) are associated with high-frequency modes, which act very similar to noise. Although, the PCA-5 and PCA-6 models are low in $R^2$ for validation data, the models are greater $80\%$ accurate in terms of nRMSE. 
We also evaluated the accuracy of combination of all the PCA models to predict the vector profiles (absolute pressure and swirl angle). We estimate the error on vector profiles by calculating the maximum percentage deviation between predicted and true vector field for a given design. The histogram of the maximum percentage deviation for the validation data set is shown in Figure \ref{Fig:profile_error}. The models are within the maximum deviation of $\pm 2\%$ for the absolute pressure profiles and $\pm 8\%$ for the swirl angle profiles.  }

\begin{figure}[h]
	\centering
	\includegraphics[width=0.6\textwidth]{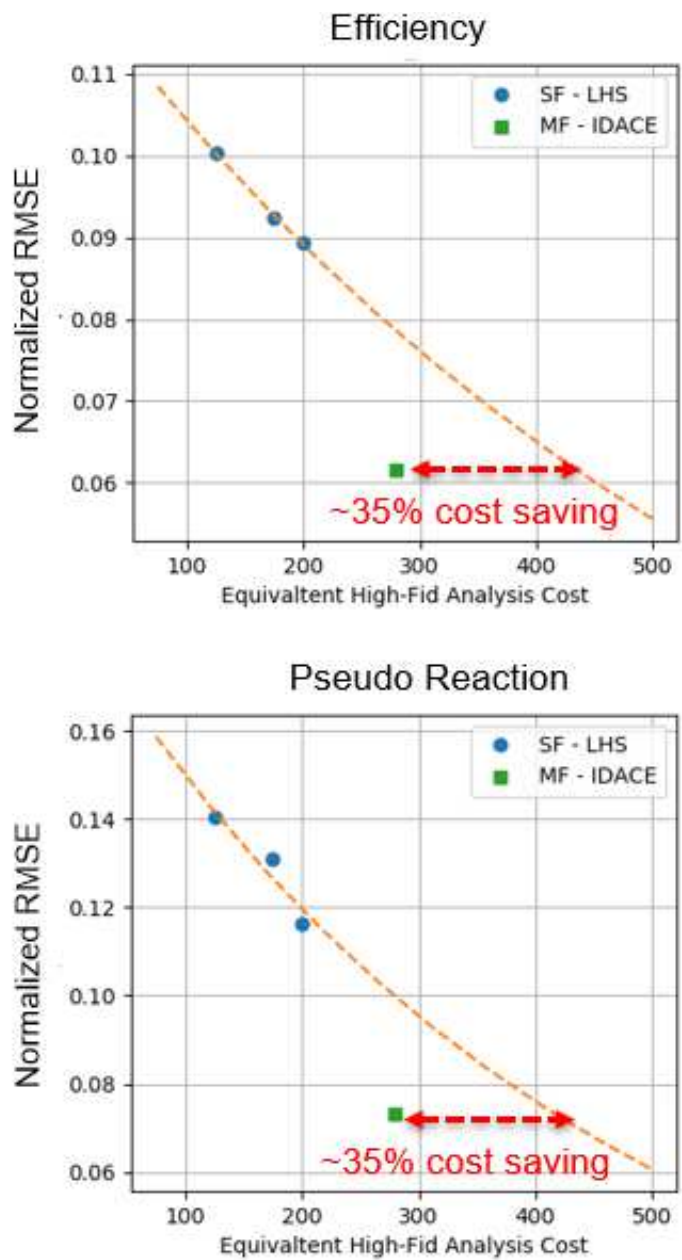}
	\caption{Cost saving in the forward modeling using MFGP and multi-fidelity adaptive sampling}
	\label{Fig:cost_saving_forwad}
\end{figure}

\highlighttext{To estimate the cost saving in the forward modeling with MFGP and multi-fidelity adaptive sampling, we also trained single-fidelity (SF) GP models with only high-fidelity data using random Latin Hypercube Sample (LHS), with sample size of 125, 175, and 200. Then we evaluate the corresponding nRMSE with the same hold-out data which was used for MFGP with multi-fidelity adaptive sampling. Based on the trend of nRMSE versus equivalent high-fidelity CFD analysis cost (No. of high-fidelity data +  (1 / cost ratio) $\times$ No. of low-fidelity data), we estimate a cost saving of at least $35\%$ in terms of computational cost associated with the forward modeling, as shown in Figure \ref{Fig:cost_saving_forwad}.}

\subsection*{Inverse Modeling}

In this work, there are eighty-five parameters considered as the input data for the cINN model.
The output data consists of the aerodynamic efficiency, the pseudo-reaction, and the pressure and the swirl angle profiles with a total of $202$ values for a given $3$D blade. 
In this work, we pre-process the input data such that each parameter follows a standard normal distribution with mean $0$ and variance $1$. 
We also normalize the output data between zero and one. 
As mentioned earlier, we develop the inverse surrogate model that maps the noisy observations to the input data. 
The inverse mapping is confined to a broad prior distribution of the input data with which we train our cINN model.

\begin{figure*}[h]
	\centering
	\includegraphics[width=1\textwidth]{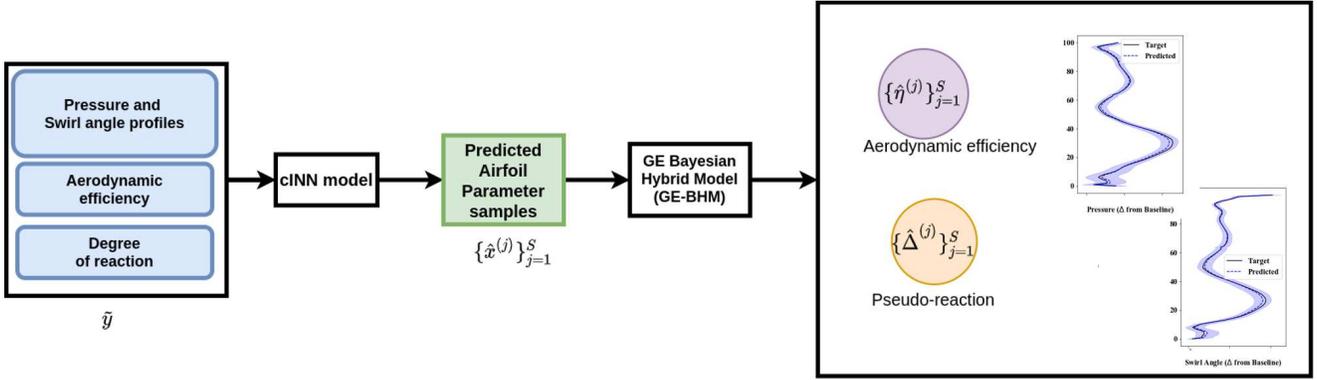}
	\caption{Framework for post-processing the samples generated from the cINN model using the GE Bayesian Hybrid Modeling. Once the airfoil parameters for each S samples are predicted, we propagate these design samples through the BHM model to generate the pressure and swirl angle profiles, aerodynamic-efficiency and pseudo-reaction.}
	\label{Fig:post}
\end{figure*}

Once the airfoil parameters for each of the samples $S$ are predicted using Algorithm~\ref{algo_test}, we propagate these design samples through the forward surrogate model (GEBHM) to generate the outputs: the aerodynamic efficiency, the pseudo-reaction, the full pressure and the swirl angle profiles, as shown in Figure ~\ref{Fig:post}. 
Similar to forward mode, we consider $R^2$ and nRMSE to evaluate the developed inverse surrogate model on the test data $\{\bm{x}^{i},\tilde{\bm{y}}^{i}\}_{i=1}^{T}$, where $T$ is the total number of test data from the forward surrogate model that was unseen when training the cINN model. 
Note that once our inverse surrogate model is trained, we can solve many inverse problems with different observations. 
Here, we test our model with different observation data.

\begin{figure}[H]
	\centering
	\includegraphics[scale=0.38]{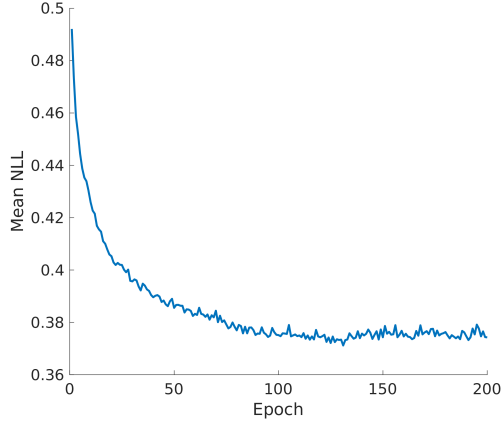}
	\caption{Mean NLL loss during training cINN for 3D blade.}
	\label{Fig:meanNLL}
\end{figure}

We train the inverse surrogate model with $N = 60000$ training data ($\mathcal{D}$) that was generated from the GEBHM. 
We train the cINN model using the Adam optimizer~\cite{kingma2014adam} with the learning rate of $1e-5$, weight decay of $5e-7$ and mini-batch size of $16$. 
The invertible network consists of $L = 8$ invertible blocks and in each invertible block, the scale $s(\cdot)$ and the shift $t(\cdot)$ networks are  fully connected layers which consists of $3$ hidden layers with $256$, $512$ and $256$ hidden units respectively and employ dropout with a keep probability of $0.2$. 
The conditioning network consists of fully connected network which consist of $4$ hidden layers with $400$, $512$, $640$, and $896$ hidden units, respectively. 
As mentioned earlier, both the conditioning network and the invertible network are trained in an end-to-end fashion.  
In this work, we train our model for $200$ epochs and use a learning rate scheduler which drops 10 times on the plateau of the mean negative log-likelihood. 
We show the mean NLL error during training for 200 epochs in Figure~\ref{Fig:meanNLL}. 
We observe that our inverse surrogate model converges at $120$ epochs, and the learning rate scheduler drops the learning rate at epoch number $70$. 

In this work, we propagate the cINN samples $\{\hat{\bm{x}}^{(j)}\}_{j=1}^{S}$ through the forward surrogate model to generate the aerodynamic efficiency, the pseudo-reaction, and the pressure and swirl angle profiles. 
Here, we consider $100$ test data that were unseen during training the cINN model. 

\begin{figure}[h!]
	\begin{minipage}[b]{0.95\linewidth}
		\centering
		\includegraphics[width=0.45\linewidth]{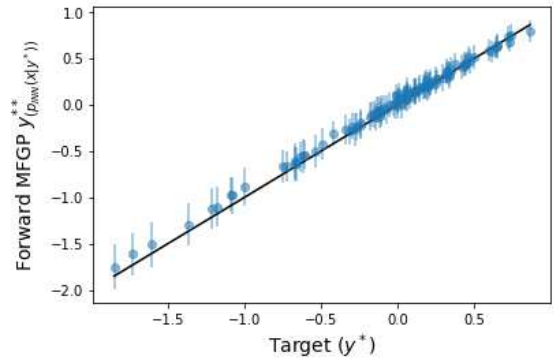} \\
		{(a) Aerodynamic efficiency}  
	\end{minipage}

	\begin{minipage}[b]{0.95\linewidth}
		\centering
		\includegraphics[width=0.45\linewidth]{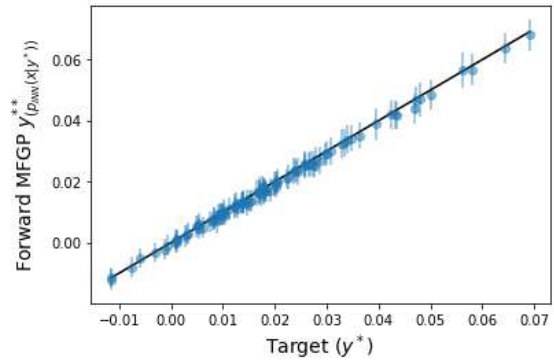}\\
		{(b) Pseudo-reaction}  
	\end{minipage}
	
	\caption{The ground truth vs. the forward BHM solution for
		the cINN model generated samples: (a) aerodynamic efficiency, and  (b) pseudo-reaction. The blue dot shows the mean of the predicted objectives, and the blue line shows the spread ($\mu \pm \sigma$) for each test data.}
	\label{fig:plot1}
\end{figure}

The ground truth vs. the forward GEBHM surrogate solution for the samples generated using the cINN model is shown in Fig.~\ref{fig:plot1}(a) for the aerodynamic efficiency, and Fig.~\ref{fig:plot1}(b) for pseudo-reaction. 
In each subplot, we show the mean $\mu$ of the predictions for the objectives mentioned above and the spread ($\mu \pm \sigma$). For both the objectives, we observe that in most of the test data, the mean $\mu$ of the predicted samples coincides with the ground truth values with a reasonable spread. 
The normalized root mean squared error and $R^{2}$ score for the aerodynamic efficiency, and the pseudo-reaction are tabulated in Table~\ref{tab:result}, and we see that the model can predict well for both the objectives with the accuracy above $95\%$.

\begin{figure}[h!]
	\begin{minipage}[b]{0.5\linewidth}
		\centering
		\includegraphics[width=0.65\linewidth]{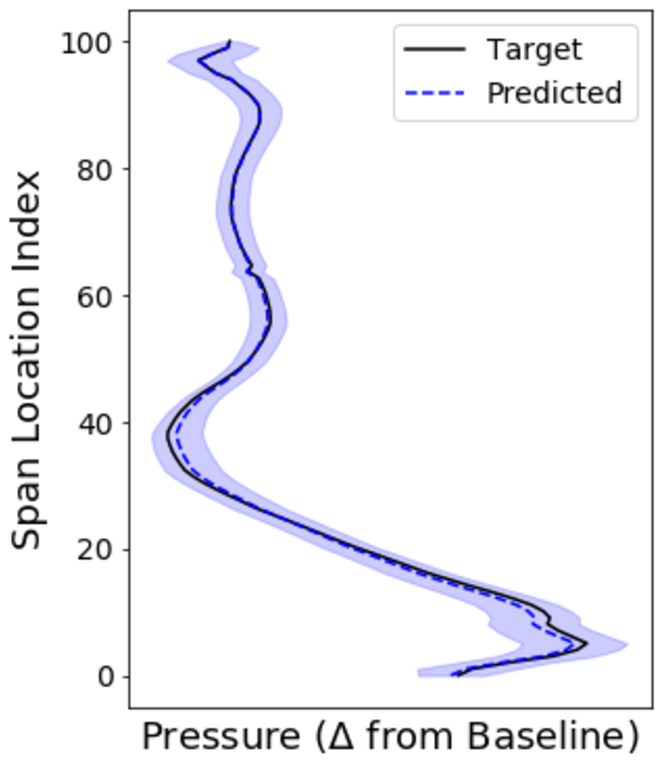} \\
		{(a)}  
	\end{minipage}
	\begin{minipage}[b]{0.5\linewidth}
		\centering
		\includegraphics[width=0.65\linewidth]{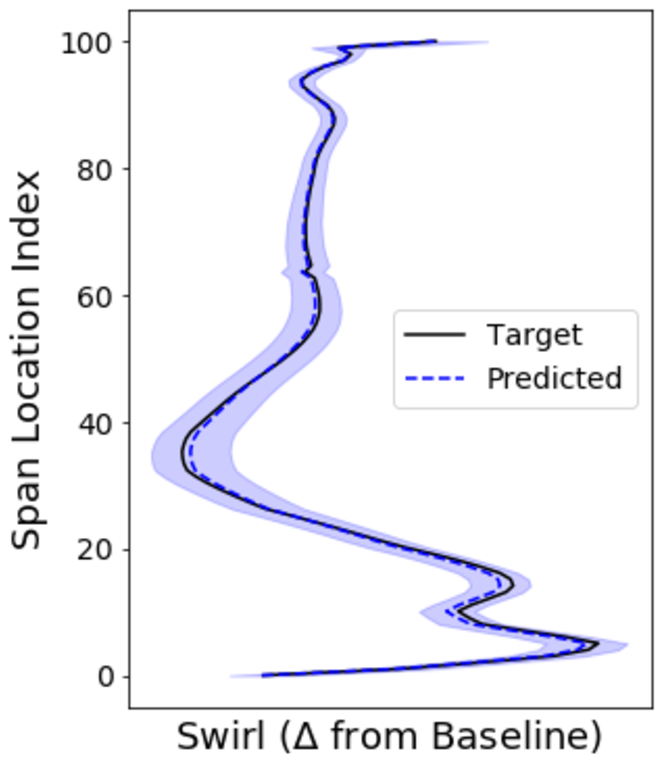} \\
		{(b) }  
	\end{minipage}
	\caption{The predicted pressure and swirl angle profile mean and the spread ($95\%$ confidence interval) along with the ground truth test data $\bm{y}^{\star}$. The black curve is the target; the blue curve is the mean of the predicted (a) pressure profile
		and (b) swirl angle profile, and the blue shaded area is $95\%$ the confidence interval.}
	\label{fig:profile}
\end{figure}

\begin{table}[h]
	\caption{Metrics for our proposed framework: Normalized root mean squared error (nRMSE) and $R^{2}$ score for the aerodynamic efficiency and the pseudo-reaction.} 
	\centering
	\begin{tabular}{l|l|l|}
		\cline{2-3}
		& $R^{2}$ & \textbf{nRMSE}  \\ \hline \hline
		\multicolumn{1}{|l|}{Efficiency} & 0.990   & 0.019  \\ \hline
		\multicolumn{1}{|l|}{Pseudo Reaction}   & 0.995   & 0.0124  \\ \hline
	\end{tabular}
	\label{tab:result}
\end{table}
In Fig.~\ref{fig:profile}, we show the predicted pressure and swirl angle profile mean and the spread ($95\%$ confidence interval) for a test data $\bm{y}^{\star}$ at the downstream of the rotor. 
We observe that the mean of the predicted pressure profile in Fig.~\ref{fig:profile} (a) and the swirl angle profile shown in Fig.~\ref{fig:profile} (b) almost coincides with the test data pressure (in Fig.~\ref{fig:profile} (a)) and swirl angle profile (in Fig.~\ref{fig:profile} (b)) respectively with a reasonable spread shown in blue shaded area. 

\subsection*{Overall CFD Validation}

To validate the 3D airfoil inverse modeling, a set of target points is provided as input to the PMI framework. 
The objectives are expressed as the target efficiency and the pseudo-reaction, together with rotor exit profiles as vectors, to retrieve a turbine design or a family of feasible designs. 
The selection of the objectives is intended to replicate a turbine design process, where a pseudo-reaction would be selected based on 1D design, and then the turbine efficiency would be targeted or optimized. 
Additionally, rotor exit flow characteristics are included as radial profiles of total pressure and swirl to account for the effect of the rotor design on the integrated system performance.

The process returns a distribution of design parameters, and five samples are selected for each target case to execute the CFD modeling and validate the objective values. 
In this process, the designs can be down selected based on additional constraints of geometrical properties. 
For this validation problem, the cross-sectional area, and the maximum thickness at section 5 (close to midspan) are selected. 
The practical application of this feature is to retrieve a design able to ensure a certain performance, while also maintaining mechanical properties within a specified range.
The designs, assessed through CFD, are in the range of desired efficiency with an uncertainty of approximately 0.2\% points. 
The uncertainty decreases towards the region of high efficiency, which is a desirable feature for design applications where optimal designs are targeted. 
The uncertainty in the prediction of pseudo-reaction as delta from the baseline value is approximately 0.01 over the design space.

\begin{figure*}[t]
	\begin{center}
		\setlength{\unitlength}{0.012500in}
		\includegraphics[width=0.9\textwidth]{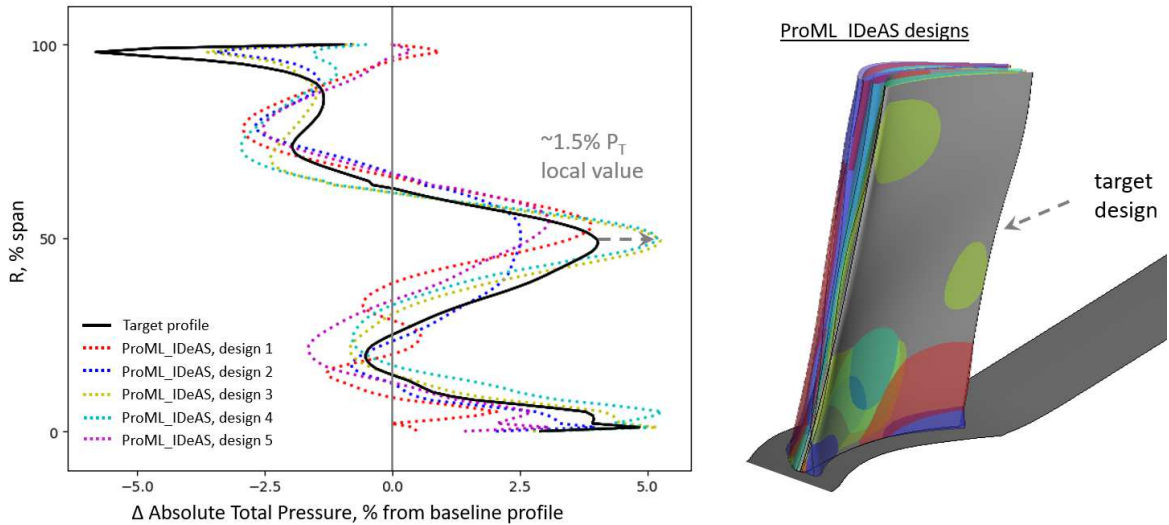}
	\end{center}
	\caption{Distribution of ideal Mach number along the airfoil for a given combination of target aerodynamic efficiency and reaction - CASE 30.}
	\label{fig:3D_validation_CASE30} 
\end{figure*}

\begin{figure}[t]
	\begin{center}
		\setlength{\unitlength}{0.012500in}
		\includegraphics[width=0.5\textwidth]{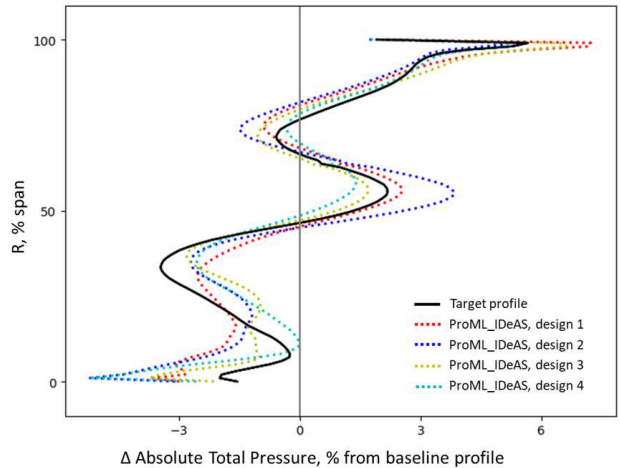}
	\end{center}
	\caption{Distribution of ideal Mach number along the airfoil for a given combination of target aerodynamic efficiency and reaction - CASE 25.}
	\label{fig:3D_validation_CASE25} 
\end{figure}

A detailed post-processing is performed on a selection of these validation cases to analyze in detail rotor performance and the exit profiles. Figure \ref{fig:3D_validation_CASE30} and Figure \ref{fig:3D_validation_CASE25} show the target profiles for two of the validation cases, and the profiles retrieved from CFD for the cINN-based designs. 
Each figure shows the target profile as a solid black line; the target geometry also provides the efficiency and reaction requested. 
The airfoils designed by the cINN are added to the charts in colored dashed lines. 
These airfoils all perform similarly to the baseline within uncertainty. 
All the profiles are shown as variations from a reference profile, calculated on the baseline design.

Two validation cases are shown in Figure \ref{fig:3D_validation_CASE30} and Figure \ref{fig:3D_validation_CASE25}. 
These target profiles are selected to verify the cINN capabilities because the profiles are skewed towards the hub and the tip respectively. 

\highlighttext{The two figures show radial profiles of absolute total pressure at the exit or the rotor, at a station downstream of the trailing edge. The location where the profiles are extracted is sketched in \ref{fig:CFD_profiles}. The quantities are averaged in the pitchwise direction, so the profiles represent variation of flow quantities along the rotor span. On the abscissa, the total pressure is reported as difference from a reference profile, as percentage of the baseline profile itself. The ordinate shows the radial location, in percentage of the rotor span. The zero indicate the rotor hub, while the one hundred is located at the turbine shroud.}

The case in Figure \ref{fig:3D_validation_CASE30} has a profile below zero for most of the upper span, meaning that the target pressure is lower than the baseline pressure. 
At midspan and blade root, there is an increase of total pressure. 
\highlighttext{The target profile is the solid black line. A set of designs, proposed by the framework, are analyzed with CFD and added to the chart. These results retrieved from CFD calculations on the cINN designs, marked with dashed lines,} all show a hub pressure profile more energized than baseline, and a higher pressure at midspan. 
Analogously, in Figure \ref{fig:3D_validation_CASE30}, the target profile has higher pressure at the tip, and the framework is able to capture the trend and impose it to the retrieved geometry. 
\highlighttext{In particular the designs 2, 3 and 4 are able to follow all the desired features, at hub, midspan and shroud. The maximum deviation from target achieved by the framework is under 2\% of the local pressure value.}
\highlighttext{On the right side of Figure \ref{fig:3D_validation_CASE30}, we show some of the blades retrieved by the PMI framework. Each blade is colored following the left plot, and the target blade is the grey surface. The blades are clearly different from the original blade, which is expected because the objective is set in terms of exit profile and not geometry. This method proposes a family of alternative designs to achieve analogous aerodynamic performance.}
\highlighttext{Similarly, in Figure \ref{fig:3D_validation_CASE25}, the second validation case is reported. The target profile in this case is pushing the total pressure at the shroud, while having a weaker hub. All the designs proposed by the framework, assessed with CFD, follow the desired trend as indicated by the colored dashed lines.}

As demostraed in the validation cases, the profiles are not restricted to an optimal solution. 
The cINN is trained using the designs generated by the forward model in the entire design space in which the forward model was trained. 
So, the cINN is capable of generating design solutions of the objectives which are possible within the design the space (including the non-Pareto-optimal solutions). 
If a non-achievable target is provided to the cINN, the cINN tends to generate design solutions which can perform close to the targets. 
To check how good the solutions of cINN are, designer can also run the forward model to evaluate the performance of the cINN solutions. 
The forward MFGP model can provide a good estimate of the performance and quantify the uncertainty of cINN design solutions. Designer can use these information to pick designs for final evaluation using expensive CFD simulation.

It is also relevant to observe that the process generates a wide range of design options which ensure both performance and exit profiles. 
This feature will be key for a multi-disciplinary design, when certain design options may not be feasible in other disciplines. 
Having multiple design options allows the designer to retain the ability to make decisions based on additional constraints that may not have been  considered in the early stages of the design process.

\section{CONCLUSION \& DISCUSSION}
In this work, we demonstrate an "explicit" inverse aerodynamic design of industrial gas turbine blade airfoil using  the Pro-ML IDeas (PMI) framework.
In  PMI, the explicit inverse design is modeled using conditional Invertible Neural Network (cINN) which is a parameterized Bayesian bijective transform between design variables and engineering quantities of interest.
To train the cINN model efficiently, data is cheaply generated using a probabilistic surrogate model of forward process.
The surrogate of the forward process is built using a MFGP with adaptive sampling, which allows us to intelligently use limited data to build an accurate model.

For the inverse design modeling of 3D last stage blade, we used only $231$ high-fidelity fine-mesh and $249$ low-fidelity coarse-mesh CFD data to build an accurate forward MFGP model with accuracy of more than 90\% for both the scalar and vector objectives.
The training of final forward MFGP with 85 input dimension model takes around 2 to 3 hours for each objective in a personal workstation without using parallel MCMC capability. In future, the trained MFGP model can be used for transfer learning for new creating new models for different turbine or operating conditions.
The inverse model is trained using $60,000$ data points generated using the forward MFGP model. 
Based on our evaluation, the cINN trained more than 98\% accurate for the scalar objectives with respect to the forward MFGP model.
The overall CFD validation results shows the inverse design generated by INN are 0.2\% points range of target efficiency and within an uncertainty of 0.01 for pseudo-reaction.
In terms of vector objectives (pressure profiles), the design generated by the framework have a maximum deviation of 2\% from target profiles.

Overall, the main computational cost associated with the PMI framework is linked with the cost and the amount of data generated to build an accurate forward generative model. Using a MFGP with adaptive sampling approach, resulted in savings of over $35\%$, in computational cost, over traditional approaches such as building GP models with one-shot design using Latin Hypercube Sampling.
Training a cINN model for the blade aerodynamics problem using extensive and cheap data from the forward model takes an order of few hours of wall clock time.
However, an inverse design query using the trained model to generate samples of design for a given target performance takes less than a second. 
In comparisons, other approaches such as Bayesian inversion approaches will take order of hours for each inversion query for the demonstrated problem.
Therefore by using PMI framework, the savings in computational cost increase linearly with an increase in the number of inverse design queries.

As discussed in previous section, the cINN can be queried for both Pareto-optimal and non-Pareto optimal solutions.  
Although if target performance associate with Pareto-front is not known upfront, finding the Pareto-optimal design can be tricky and challenging. 
Alternatively, since, querying the cINN and forward models are very cheap ($O(milliseconds)$), designer can query multiple combination of extreme target performance (close to possible Pareto-front), find the design solution using cINN and quantify the uncertainty associated with the design performance using the forward model.

The current framework mainly focuses on “equality” type objectives (and constraints). However, the inverse design with inequality can also be handled by the invertible network by querying the constraints values which satisfies the constraint. 
For example, consider a design problem with target objective $y(\bm{x})=y_{target}$ and inequality constraint $g(\bm{x}) \le 0$. 
One can build an cINN with y and g as objectives. With the trained cINN, multiple queries to cINN can be made with target satisfying the constrains, such $(y_{target},g_1 ),(y_{target},g_2),(y_{target},g_3),...$ etc., where $g_i \le 0$. 
These queries will generate multiple designs which satisfies the design target as well as inequality constraint, from which the designer can pick solution based on their expertise and preference. 

The main challenge of training cINN models are the number of data required for training such models. 
However, from the industrial standpoint, the data associated with design are generally sparce and expensive to generate. 
Our main contribution of this work came through overcoming this obstacle by developing the 2-step PMI framework for training cINN for industrial challenging high-dimensional problems. 
In the first step of PMI framework, a fast and accurate forward model is developed using multi-fidelity data in an adaptive manner to decrease the data requirement. 
In the second step, the cINN is trained using cheap and large data generated using the fast forward model.
Additionally, we also extended the real NVP with a conditioning network in this work to handle challenges associated with deep generative models such as training stability and capturing sharp features in the underlying physical process.

In this work, we demonstrated the above framework on only the aerodynamic aspect of turbine blade design. 
Directions for future work include extending the framework to the multidisciplinary and aero-mechanics aspects of design, involving structural, modal, and flutter analysis. 
Moreover, the focus of demonstrating the capability of the proposed framework is on a challenging problem on industrial gas turbine (IGT) blades, however the application of the framework is not limited to IGTs.

\begin{acknowledgment}
The information, data, or work presented herein was funded
in part by the Advanced Research Projects Agency-Energy (ARPA-E), U.S.
Department of Energy, under Award Number DE-AR0001204. 
The views and opinions of authors expressed herein do not necessarily state or reflect those of
the United States Government or any agency thereof.
\end{acknowledgment}

%
% ...

\bibliographystyle{asmems4}
\bibliography{sample}
%
%

\end{document}